\documentclass[useAMS,usenatbib]{mn2e}
\usepackage{graphicx}
\usepackage{txfonts}

\def\apj{ApJ}

\def\aap{A\&A}

\def\mnras{MNRAS}

\def\beq#1{\begin{equation}\label{#1}}
\def\eeq{\end{equation}}
\def\beqa#1{\begin{eqnarray}\label{#1}}
\def\eeqa{\end{eqnarray}}

\def\Eq#1{Eq.~(\ref{#1})}

\def\myfrac#1#2{\left(\frac{#1}{#2}\right)}
\def\mean#1{\langle{#1}\rangle}
\def\comment#1{\relax}

\title[`Off' states in X-ray pulsars]{On the nature of `off' states in 
slowly rotating low-luminosity X-ray pulsars}
\author[N. Shakura et al.] {N. Shakura
\thanks{E-mail: nikolai.shakura@gmail.com, kpostnov@gmail.com},
K. Postnov, 
L. Hjalmarsdotter\\
Sternberg Astronomical Institute, Moscow M.V. Lomonosov State University, Universitetskij pr., 13, 119992, Moscow, Russia}

\begin{document}

\date{Received ... Accepted ...}
\pagerange{\pageref{firstpage}--\pageref{lastpage}} \pubyear{2012}

\maketitle

\label{firstpage}

\begin{abstract}
We elaborate on a recently proposed model for subsonic quasi-spherical accretion onto slowly rotating pulsars, in which accretion is mediated through a hot quasi-static shell above the neutron star magnetosphere. We show that under the same external conditions, two regimes of subsonic accretion are possible, depending on if plasma cooling in the transition zone is dominated by Compton or radiative processes. We suggest that a transition from the higher luminosity Compton cooling regime to the lower luminosity radiative cooling regime can be responsible for the onset of  the `off'-states repeatedly observed in several low luminosity slowly accreting pulsars, such as Vela X-1, GX 301-2 and 4U 1907+09. We further suggest that the triggering of the transition may be due to a switch in the X-ray beam pattern in response to a change in the optical depth in the accretion column with changing luminosity. 
\end{abstract}


\section{Introduction}
\label{intro}

Quasi-spherical accretion from stellar winds onto magnetized neutron stars in binary systems may proceed in 
two physically distinct ways. The gravitationally captured stellar wind matter is  
heated behind the bow shock at the characteristic Bondi radius $R_B\approx 2GM/v_w^2$ 
where $v_w$ is the relative wind velocity. If this matter cools down on a time scale
shorter than the free-fall time $t_{ff}=R^{3/2}/\sqrt{2GM}$, 
it will fall at a supersonic velocity towards the magnetosphere and come to a halt in a shock above it.
The plasma cools down most effectively via Compton scattering of X-ray photons generated near the neutron star surface. Therefore for such supersonic (Bondi-type) accretion to take place a dense 
photon field is required. Thus, this regime of accretion is supposed to take place at high X-ray luminosities and was studied in detail, for example, in \citet{AronsLea76, Burnard_ea83}. In the Bondi accretion regime
the matter cannot pile up above the magnetosphere, and the mass accretion 
rate onto the neutron star $\dot M$ is ultimately determined by the Bondi-Hoyle-Littleton 
formula for the gravitational capture mass rate 
by a moving neutron star $\dot M\simeq \rho_w R_B^2 v_w$. 

At low or moderate X-ray luminosities the captured wind matter may 
have no time to cool down, the mass fall rate becomes subsonic, and  
a hot quasi-spherical shell is due to be formed above the
neutron star magnetosphere \citep{DaviesPringle1981}. 
At the base 
of the shell 
the plasma must cool down to some critical temperate \citep{ElsnerLamb77} in order to enter 
the magnetosphere via instabilitites, and the rate of plasma entry 
into the magnetosphere will be determined by these instabilities.
Therefore, 
the velocity of matter settling through the shell will be regulated 
by the ability of the 
plasma to enter the magnetosphere. As was shown in (\citealt{Shakura_ea12}, Paper I
hereafter), the extended quasi-static shell  
mediates the angular momentum removal from the rotating 
magnetosphere by large-scale convective motions.
The mass accretion rate onto the neutron star is determined by the density
above the magnetosphere and the mean settling velocity of matter through the shell, 
and can be very small if plasma cooling above the magnetosphere is inefficient.

It was shown in Paper I that a settling regime of accretion onto 
slowly rotating neutron stars can be established 
for X-ray luminosities $L_x\lesssim L*\approx 4\times 10^{36}$~erg/s 
corresponding to accretion rates 
$\dot M\lesssim \dot M*\simeq 4\times 10^{16}$~g/s. At higher accretion
rates, a free-fall gap 
above the magnetosphere appears in the flow due to rapid Compton cooling, and accretion 
becomes highly non-stationary. 

The model of subsonic quasi-spherical settling accretion presented in Paper I 
is based on generic properties of wind accretion onto magnetized
neutron stars with moderate and low X-ray luminosities, and can be applied to a variety of sources. 
For example, in Paper I we applied this model to observations of
the slowly rotating X-ray pulsars Vela X-1 and GX 301-2 in, both in high-mass X-ray
binaries and spinning at an
equilibrium period, as well as to the steadily spinning-down X-ray pulsar GX 1+4 in a symbiotic X-ray binary, in which 
negative spin-down-luminosity correlations are observed (see also \citealt{GonzalezGalan_ea12}). 
The model was also successfully used by \cite{Lutovinov_ea12}
to describe the observed spin-luminosity correlations in the slowly rotating
low-luminosity X-ray pulsar X Per in 
an Be/X-ray binary as well as the observed long period in 
the Be/X-ray binary SXP 1062 \citep{PopovTurolla12}. The model was also recently used 
in population synthesis studies \citep{ChashkinaPopov12, Lue_ea12}.

In this paper we further develop the model by focusing on two 
different regimes (within our subsonic model) of plasma entering the neutron star magnetosphere. In Section 2 we show that 
under the same external conditions in the wind ($\rho_w, v_w$), which largely determine the 
plasma density distribution in the quasi-spherical hot shell, 
subsonic accretion through the shell can occur in two distinct 
regimes depending on
the characteristic cooling time of the plasma above the Alfv\'en surface: the
Compton (shorter time scale, higher luminosity) and radiative (longer time scale, lower luminosity) regime. 
The plasma cooling time $t_{cool}$  determines 
the mean velocity of matter falling through the transition zone. In this zone,
the Rayleigh-Taylor instability, which allows plasma to enter the magnetosphere, develops. 
This velocity is inversely proportional to the plasma cooling time, $u_R\sim
t_{cool}^{-1/3}$, and eventually determines 
the mass accretion rate through the magnetosphere $\dot M$ onto the neutron star.
Therefore, in the same source accretion in both a 
higher luminosity (Compton cooling dominated) and
lower luminosity (radiative cooling dominated) regime is possible. 

We identify these higher and lower luminosity regimes, respectively,  
with `normal' luminosity levels and the occasional `off'  
states as observed in some X-ray pulsars as e.g. Vela X-1, GX 301-2 and 4U1907+05, which are discussed in Section 3. 

In Section 4 we show that the transition from the higher luminosity to lower luminosity
regime may be related to a sudden decrease in X-ray photon energy density 
in the equatorial region of 
the magnetosphere, which is most favourable for the plasma to enter due 
to the Rayleigh-Taylor instability \citep{AronsLea76}. Such a decrease  
may be the result of a  
change in the X-ray beam pattern from the accretion column (or `mound' above the polar cap) 
when the X-ray luminosity drops below some critical value $L_\dag\sim 3\times 10^{35}$~erg/s
determined by the opacity of the column relative to Thomson scattering of X-ray photons. 
Below this luminosity most of the X-ray emission escapes in a pencil beam, so most 
of the X-ray photons illuminate the magnetospheric cusp region, which is stable
for plasma entering the magnetosphere. However, plasma cooling continues on the longer radiative cooling time scale, 
which is determined only by the density and temperature above the magnetosphere, 
and the source switches into the lower luminosity regime.  
Oppositely, an increase in photon energy density in the equatorial
magnetospheric region can return the source to the higher luminosity regime.


\section{Two regimes of subsonic accretion from a quasi spherical shell}

Consider a hot quasi-spherical shell formed around a slowly
rotating neutron star magnetosphere, in which accretion proceeds subsonically (see Paper I for details). 
The shell can exist as long as the X-ray luminosity is less than $\sim 4\times 10^{36}$~erg/s, 
above which supersonic (Bondi) accretion is more likely \citep{Burnard_ea83}.

To enter the magnetosphere, the plasma in the shell must cool down from a  
high (almost virial) temperature $T$ determined by hydrostatic equilibrium 
[Eq (4) in paper I] to $T_{cr}$ \citep{ElsnerLamb77} 
\beq{30}
{\cal R}T_{cr}=\frac{1}{2(1+\gamma m_t^2)}\frac{\cos\chi}{\kappa R_A}\frac{\mu_mGM}{R_A}
\eeq
Here ${\cal R}$ is the universal gas constant, $\mu_m\approx 0.6$ is the molecular weight,
$G$ is the Newtonian gravitational constant, $M$ is the neutron star mass,  
$\kappa$ is the local curvature of the magnetosphere, $\chi$ is the angle 
between the outer normal and the radius-vector at any given point at the Alfv\'en surface, and the contribution 
of turbulent pulsations in the plasma to the total pressure is
taken into account by the factor $(1+\gamma m_t^2)$ (where $m_t$ is the turbulent Mach number
$\gamma=C_P/C_V$ is the ratio of specific heat capacities).

As was shown in Paper I, a transition zone above the 
Alfv\'en surface with radius $R_A$ is formed inside which the plasma cools down. The effective gravitational acceleration in this zone is 
\beq{}
g_{eff}=\frac{GM}{R_A^2}\cos\chi \left(1-\frac{T}{T_{cr}}\right)
\eeq 
and the mean radial velocity 
of plasma settling is $u_R=f(u)\sqrt{2GM/R_A}$. The dimensionless settling velocity 
$0\le f(u)\le 1$ is determined by the specific plasma cooling mechanism 
in this zone and is constant through the shell.  
Together with the density of 
matter near the magnetospheric boundary $\rho(R_A)$ it determines the magnetosphere 
mass loading rate through 
the mass continuity equation:
\beq{cont}
\dot M=4\pi R_A^2 \rho(R_A) f(u)\sqrt{2GM/R_A}\,.
\eeq
This plasma eventually reaches the neutron star surface and 
produces an X-ray luminosity $L_x=0.1\dot M c^2$.
Below we shall normalize the mass accretion rate 
through the magnetosphere as well as the X-ray luminosity 
to the fiducial values $\dot M_{16}=\dot M/10^{16}$~g/s and $L_{36}=L_x/10^{36}$~erg/s, respectively.

The stationary settling velocity is 
determined by the plasma cooling time $t_{cool}$ in the transition zone
(see Paper I):
\beq{fu}
f(u) \simeq \myfrac{t_{ff}}{t_{cool}}^{1/3}\cos\chi^{1/3}
\eeq
where $t_{ff}=R^{3/2}/\sqrt{2GM}$ is the characteristic free-fall time from radius $R$. The angle
$\chi$ is determined by the shape of the magnetosphere, and for the magnetospheric boundary 
parametrized in the form   
$\sim \cos\lambda^n$ (where $\lambda$ is the angle counted from the magnetospheric equator) 
$\tan \chi=n\tan\lambda$. For example, in model calculations by \cite{AronsLea76} $n\simeq 0.27$ 
in the near-equatorial zone. We see that $\cos\chi\simeq 1$ up to $\lambda\sim \pi/2$, 
so below (as in Paper I) we shall omit this factor.

\subsection{The Compton cooling regime}

As explained in detail in Paper I (Appendix C and D), in subsonic quasi-static shells above slowly rotating neutron star magnetospheres such as considered here,
the adiabaticity of the accreting matter is broken due to turbulent heating and Compton cooling. 
X-ray photons generated near the neutron star surface tend to cool down the matter in the shell via Compton scattering as long as the plasma
temperature $T>T_x$, where $T_x$ is the characteristic radiation temperature determined
by the spectral energy distribution of the X-ray radiation. 
For typical X-ray pulsars $T_x\sim 3-5$~keV. 
Cooling of the plasma at the base of the shell decreases the tempreature gradient 
and hampers convective motions. Addtional heating due to   
sheared convective motions is insignificnt (see Appendix C of Paper I).  
Therefore, the temperature in the shell 
changes with radius almost
adiabatically ${\cal R}T\sim (2/5)GM/R$, and 
the distance $R_x$ within which the plasma cools down 
by Compton scattering is
\beq{}
R_x\approx 10^{10}\hbox{cm}\myfrac{T_x}{3\hbox{keV}}^{-1}\,, 
\eeq 
which is much larger than the characteristic Alfv\'en radius $R_A\simeq 10^9$~cm.

The Compton cooling time is inversely proportional to the 
photon energy density, 
\beq{t_C}
t_C\sim R^2/L_x\,, 
\eeq 
and near the Alfv\'en surface we find
\beq{t_Cn}
t_C\approx 10 \hbox{[s]} \myfrac{R_A}{10^9 \hbox{cm}}^2 
\myfrac{L_x}{10^{36}\hbox{erg/s}}^{-1}\,.
\eeq
This estimate assumes spherical symmetry. 
Clearly, for the exact radiation density the shape of the X-ray emission  
produced in the accretion column near the neutron star surface
is important (see Sections 3 and 4), but still $L_x\sim \dot M$. 
Therefore, roughly, $f(u)_C\sim \dot M^{1/3}$, or, more precisely, 
taking into account the dependence of $R_A$ on $\dot M$ in this regime
(see Paper I)
\beq{RA_C}
R_A^C\approx 10^9\hbox{cm}\myfrac{L_x}{10^{36}\hbox{erg/s}}^{-2/11}\mu_{30}^{6/11}
\eeq
we obtain:
\beq{fu_C}
f(u)_C\approx 0.3\myfrac{L_x}{10^{36}\hbox{erg/s}}^{4/11}\mu_{30}^{-1/11}\,.
\eeq
Here $\mu_{30}=\mu/10^{30}$~G cm$^3$ is the neutron star dipole magnetic moment. As
shown in Paper I, in a spherically symmetric accretion flow with turbulence heating
and Compton cooling, when $f(u)_C\sim 0.5$ (which corresponds to 
$L_x\sim 4\times 10^{36}$~erg/s) the sonic point 
emerges above the magnetosphere, accretion becomes supersonic and is most likely  
described by the physical model studied by \cite{Burnard_ea83}.

In the Compton cooling regime a 
significant X-ray flux variability is expected. 
Suppose that the energy density of X-ray photons
in the cooling region increases. Then the settling velocity $f(u)_C\sim \dot M^{1/3}$
increases as well, leading to further increase in the radiation energy density.
Clearly, this is a non-stationary situation. The maximum accretion rate here
will be determined by the ability of the magnetosphere to engulf the total amount of mass
within the cooling region $R<R_x$ over the free-fall time $t_{ff}(R_x)$:
\beq{}
\dot M_{max}\sim \frac{\Delta M(R_x)}{t_{ff}(R_x)}\,.
\eeq
Taking into account the hydrostatic density distribution $\rho(R) \sim R^{-3/2}$, we
find 
\beq{}
\dot M_{max}\sim 4\pi R_A^2 \rho(R_A) \sqrt{2GM/R_A}\cdot \frac{2}{3}\left(1-\myfrac{R_A}{R_x}^{3/2}\right)\,.
\eeq
Eliminating density via the mass continuity equation expressed for the 
mean mass accretion rate $\mean{\dot M}$, we obtain 
\beq{var}
\frac{\dot M_{max}}{\mean{\dot M_C}}\sim \frac{1}{f(u)_C}\approx 2
\eeq
We stress that this variability occurs on the time scale $\sim t_{ff}(R_x)$ (about a few 
hundreds of seconds) and 
will be present even if the density 
near the magnetosphere is constant. Clearly, variable external conditions would add to
the X-ray flux variability on its own time-scale. 
Also note that this instability does not destruct the shell as a whole.
The reason for this is that above the radius $R\sim R_x$ Compton heating is effective, so an increase in 
the X-ray luminosity would tend to prohibit accretion from the upper layers of the shell.

\subsection{The radiative cooling regime}

In the absence of a dense photon field, 
at the characteristic temperatures near the magnetosphere $T\sim 50$-keV and higher,
plasma cooling is essentially due to radiative losses (bremsstrahlung), and the plasma cooling time is 
$t_{rad}\sim \sqrt{T}/\rho$. Making use of the continuity equation (\ref{cont})
and the temperature distribution in the shell $T\sim 1/R$, we obtain 
\beq{t_rad}
t_{rad}\sim R \dot M^{-1} f(u)\,.
\eeq
Note that, unlike the Compton cooling time (\ref{t_C}), the radiative cooling time is 
actually independent of $\dot M$ (remember that $\dot M\sim f(u)$ in the subsonic accretion
regime!). 
Numerically, near the magnetosphere we have 
\beq{t_radn}
t_{rad}\approx 1000 \hbox{[s]} \myfrac{R_A}{10^9 \hbox{cm}}\myfrac{L_x}{10^{36}\hbox{erg/s}}^{-1}\myfrac{f(u)}{0.3}\,.
\eeq

Following the method described in Section 3 of Paper I, we find the 
mean radial velocity of matter entering the neutron star magnetosphere in the 
near-equatorial region,
\beq{fu_rad0}
f(u)_{rad}=\zeta^{2/3}\left[\frac{1}{6}\frac{t_{ff}}{t_{rad}}\right]^{1/3}\,,
\eeq
similar to the expression for $f(u)$ in the Compton cooling region \Eq{fu_C}. 
Here $\zeta\le 1$ is a numerical factor describing the radial extension of 
the transition zone $\zeta R_A$. Using 
the expression for the Alfv\'en radius through $f(u)$, we calculate
the dimensionless settling velocity:  
\beq{fu_rad1}
f(u)_{rad}\approx 0.1 \zeta^{14/27} L_{36}^{6/27}\mu_{30}^{2/27}
\eeq 
and the Alfv\'en radius:
\beq{RA_rad}
R_A^{rad}\simeq 10^9\hbox{[cm]} L_{36}^{-6/27}\mu_{30}^{16/27}
\eeq
(in the numerical estimates we assume a monoatomic gas with adiabatic index 
$\gamma=5/3$ and turbulent Mach number $m_t=1$ in the shell). 
The obtained expression for the dimensionless settling velocity of matter 
\Eq{fu_rad1} in the radative cooling regime clearly shows that here accretion proceeds
much less effectively than in the Compton cooling regime (cf. with \Eq{fu_C}).   

Unlike in the Compton cooling regime, 
in the radiative cooling regime there is no instability leading to an increase of the mass
accretion rate as the luminosity increases (due to the long characteristic
cooling time), 
and accretion here is therefore expected to occur more quietly
under the same external conditions.

\begin{figure*}
\includegraphics[width=\textwidth]{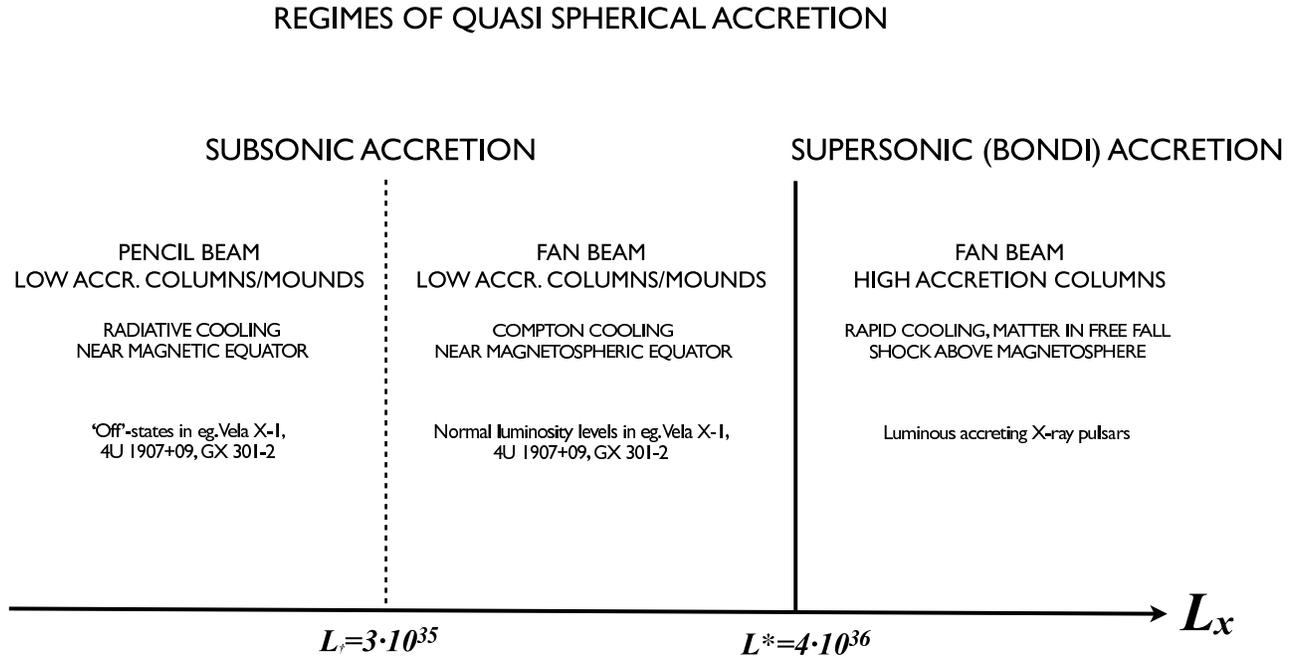}
\caption{Scheme of the different regimes of quasi-spherical accretion including supersonic (Bondi) accretion and subsonic (our model) accretion.}
\label{f:accregimes}
\end{figure*}

In Fig. \ref{f:accregimes} we summarize the different regimes of quasi-spherical accretion.

\section{`Off' states in X-ray pulsars}

\begin{figure*}
\includegraphics[width=\textwidth]{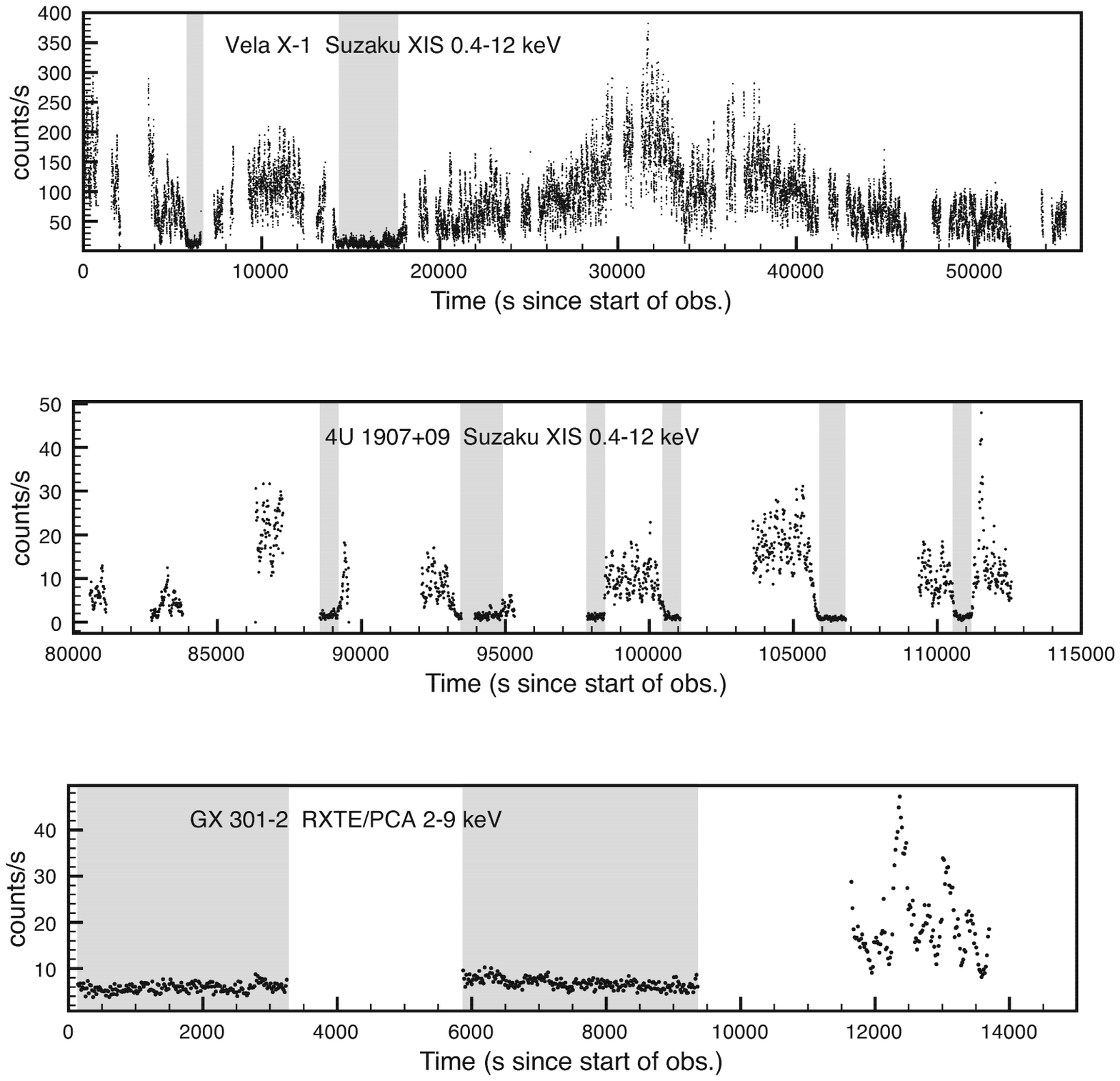}
\caption{Example lightcurves of Vela X-1, 4U 1907+09 and GX 301-2 (\textit{HEASARCH} archive data). Shaded areas mark observed off-states. Note that the timescale is different in the different panels.}
\label{f:lightcurves}
\end{figure*}

So called `off' states have been observed in several slowly rotating low luminosity pulsars such as Vela X-1 \citep{Inoue_ea84, Kreykenbohm_ea99, Kreykenbohm_ea08, Doroshenko_ea11}, 
GX 301-2 \citep{Gogus_ea11} and 4U 1907+09 \citep{intZand_ea97, Sahiner_ea12}). 
These states are characterized by a sudden, most often without any prior indication, drop in X-ray flux down to $1-10 \%$ of normal levels, lasting typically for a few minutes. A few examples of lightcurves including off-states are shown in Fig.\ref{f:lightcurves}.

It seems to bee fairly well established that the off states can not be due simply to increased absorption along the line of sight. Their occurrence is not correlated with increased $N_H$ (\citealt{Furst_ea11}, \citealt{Sahiner_ea12} but see also \citealt{Kretschmar_ea99} for a another type of intensity dips in Vela X-1, most probably caused by dense blobs in the wind), the timescale of their onset are too short (e.g. \citealt{Kreykenbohm_ea08})
and spectral studies show a softening of the X-ray spectrum during the off state \citep{Gogus_ea11, Doroshenko_ea11}, contrary to what expected had the decreased flux levels been caused by increased absorption. 
Failure by early observations with instruments like \textit{RXTE}/PCA to detect pulsations during the 
off states have seamed to suggest that the sources were instead simply turned off due to 
the sudden cessation of accretion. The popular view is that the cause of this may be large 
density variations in the stellar wind possibly combined with the onset of the propeller regime (see e.g. \citealt{Kreykenbohm_ea08}). 

Recent observations with the more sensitive instruments onboard Suzaku of Vela X-1  
\citep{Doroshenko_ea11}, however, show that although dropping in luminosity by a factor of ca 20 
the source is clearly detected with a pulse period equal to that observed at normal flux levels. 
This suggests that rather than cessation of accretion, the off-states may be better explained 
by a transition to a different, less effective, accretion regime. We suggest that the onset 
of the off state in these sources marks a transition from the Compton cooling dominated 
to the radiative cooling dominated regime as described above.

\section{Transitions between the Compton and radiative cooling regimes -- a change in beam pattern?}

In this section we will discuss how transitions between 
the two regimes of plasma entering the magnetosphere may be triggered. 

A decrease in the X-ray photon energy density in the transition zone decreases
the Compton cooling efficiency, but the Compton cooling time remains much
shorter than the radiative cooling time down to very small luminosities
(see \Eq{t_Cn} and \Eq{t_radn}). Therefore, in the spherically symmetric case, a transition between the two regimes would
require an almost complete switch-off of the Compton cooling in the
equatorial magnetospheric region. In the more realistic 
non-spherical case, the Compton cooling time can become comparable 
to the radiation cooling time when the X-ray beam pattern changes 
with decreasing X-ray luminosity from a fan beam to a pencil beam, and the 
equatorial X-ray flux is reduced by a factor of a few. Additionally, hardening of the pulsed X-ray 
flux with decreasing X-ray luminosity, which is observed in low-luminosity X-ray pulsars \citep{Klochkov_ea11}, increases $T_x$ and decreases the specific Compton cooling rate of the
plasma $\propto (T-T_x)/t_C$, thus making Compton cooling less efficient. 
Such transitions have been observed in transient X-ray pulsars 
(see, e.g., \citep{Parmar_ea89}).
The radiation density in the pencil beam
cools down the plasma predominantly in the magnetospheric cusp region, but because of 
the stronger magnetic line curvature \citep{AronsLea76} the plasma entry rate  
through the cusp will be insignificant. Still, the plasma 
continues to enter the magnetosphere via instabilities 
in the equatorial zone, but at 
a lower rate determined by the longer radiative cooling timescale.  

The mass accretion rate in the radiative cooling regime 
will be determined by the plasma density by the time  
Compton cooling switches off in the magnetospheric equator region.
This occurs at some X-ray luminosity $L_x\lesssim L_{cr}$. 


In the case of a strong neutron star magnetic field ($\gtrsim 10^{12}$~G)
most of the thermal X-ray photons produced in the energy release zone 
are produced with the ordinary (O) polarization mode, and the number of extraordinary 
(X) photons is small. Depending on the plasma density, the vacuum polarization effects
leading to conversion of O-photons into X-photons can be important for 
photons with energies between $\hbar \omega_v\approx 12 (n_e/10^{22}\hbox{cm}^{-3})^{1/2}
(B/10^{12}\hbox{G})^{-1}$~keV and the cyclotron resonance energy 
$E_c=\hbar\omega_c\approx 11.6 (B/10^{12}\hbox{G})$~keV 
\citep{Ventura_ea79, MeszarosNagel85, HardingLai07}. 
From equating $\hbar\omega_v=\hbar\omega_c$, 
we find the critical electron number density  
\beq{}
n_e\simeq 7.6\times 10^{21}[\hbox{cm}^{-3}] \myfrac{B}{10^{12}\hbox{G}}^4
\eeq
below which vacuum polarization effects on the photon mode propagation are significant. 
Since $n_e\sim \dot M/A\sim \dot M R_A$ (here $A\sim \pi R_{NS}^2 \theta_c^2\sim R_{NS}^3/R_A$ is 
the effective area of the NS surface onto which accretion proceeds) increases with mass accretion rate,
in bright X-ray pulsars with $L_x\gtrsim 10^{37}$~erg/s vacuum polarization 
effects are not expected to play a significant role, and 
the X-ray emission beam should consist mostly of O-photons.

O-photons with energies below the cyclotron resonance energy $E_c$ 
propagate along the magnetic field with a scattering cross-section $\sigma_{||}
\approx \sigma_T \sin^2\vartheta$ where $\vartheta$ is the angle between the
photon wave vector and the magnetic field \citep{HardingLai07}. 
They form a pencil beam with a 
characteristic opening angle $\theta_p\sim 1/\sqrt{\tau_T}$, where
$\tau_T$ is the optical depth in the energy release zone \citep{Basko76, Dolginov_ea79}.
These photons exert a low force on the accreting matter, 
and if the number of hard photons with energies
above $E_c$ (where $\sigma_{||}\approx \sigma_T$) is small, no high accretion column 
will be formed. 
Apparently, this is the case in some luminous X-ray pulsars (Cen X-3, Her X-1) with 
$L_x\sim 10^{37}$~erg/s.


At lower densities (corresponding to the low mass accretion rates and low X-ray luminosities
we are considering here) 
or higher magnetic fields, 
vacuum polarization is significant, and O-photons can be converted into X-photons. 
The scattering cross-section of such photons is 
$\sigma_\perp\simeq \sigma_T (E/E_c)^2$ and
is independent of the angle $\vartheta$. Therefore, they are expected to 
produce a more spherically 
symmetric (but still not fan-like) beam. Again, no high accretion column is expected to form if the number of hard photons with $E>E_c$ is small. 


Additionally, photons must be scattered in the accretion flow
above the polar cap region. This scattering does not significantly affect the pencil beam formed 
by O-photons in bright pulsars because of the $\sin^2\vartheta$ dependence of the 
scattering cross-section. In low-luminosity pulsars, in which vacuum polarization 
photon mode conversion occurs, 
Thomson scattering would tend to form a fan beam at all 
photon energies.
Therefore, the transition from fan to pencil beam in low-luminosity X-ray 
pulsars without high
columns does not occur until the optical depth in the accretion flow
above the polar cap becomes less than one. 
The optical depth in the accretion flow in the direction 
normal to the neutron star surface from the radial distance $r_6=r/10^6$~cm 
is estimated to be \citep{Lamb_ea73}
\beq{tau_v}
\tau_v\simeq 3 \myfrac{R_A}{10^9\hbox{cm}}^{1/2}\dot M_{16} r_6^{-3/2}
\eeq
(here the neutron star mass is assumed to be 1.5 $M_\odot$ and the NS radius $R_{NS}=10^6$~cm).
Taking into account the dependence of the Alfv\'en radius on $\dot M$ and $\mu$ (\ref{RA_C}), 
we see from this estimate that the X-ray diagram change is expected 
to occur at $\tau_v<1$, corresponding to an X-ray luminosity of  
\beq{L_cross}
L_{\dag}\sim 3\times 10^{35}[\hbox{erg/s}]\mu_{30}^{-3/10}\,. 
\eeq

The decrease in radiation energy density in the magnetospheric
equatorial zone due to the X-ray beam pattern
change leads to an increase of the Compton cooling time and hence 
triggers a transition
to the lower luminosity regime and the source enters the `off' state.   
From the mass continuity equation we then find the luminosity ratio:
\beq{}
\frac{L_{x,rad}}{L_{\dag}}= \frac{f(u)_{rad}}{f(u)_{C}}\sim\myfrac{t_C}{t_{rad}}^{1/3}\,.
\eeq 
Substituting expressions for $f(u)$'s \Eq{fu_C} and \Eq{fu_rad1} taken at $L_x=L_\dag$, 
we find the X-ray luminosity in the 'off' state:
\beq{Lxrad}
L_{x,rad}\approx 10^{35}[\hbox{erg/s}]\mu_{30}^{7/33}\,.
\eeq

We stress here that this X-ray luminosity is derived for the case of a shell 
around the neutron star magnetosphere in the radiation cooling regime. Lower 
X-ray luminosities can be realized only if the density of matter near the 
magnetosphere $\rho(R_A)$ (which is determined by the density of gravitationally 
captured stellar wind behind the shock at the Bondi radius $\rho(R_B)$) 
turns out to be smaller
than the value that provides the minimum X-ray luminosity $\sim L_\dag$ of the source, 
which is 
required for effective Compton cooling to operate in the equatorial region of the NS
magnetosphere. 


The return from radiative cooling dominated accretion back to the Compton cooling
dominated regime can take place, for example, due to 
a density increase above the magnetosphere, leading to an increase of the  
mass accretion rate. In turn, this leads to 
growth of the vertical optical depth of the accretion column, 
disappearance of the beam 
and enhancement of the lateral X-ray emission. The radiative energy density 
in the equatorial magnetospheric region strongly increases, Compton cooling
resumes, and the source comes back to normal luminosity levels. 

\subsection{The changing beam pattern in Vela X-1}

\begin{figure*}
\includegraphics{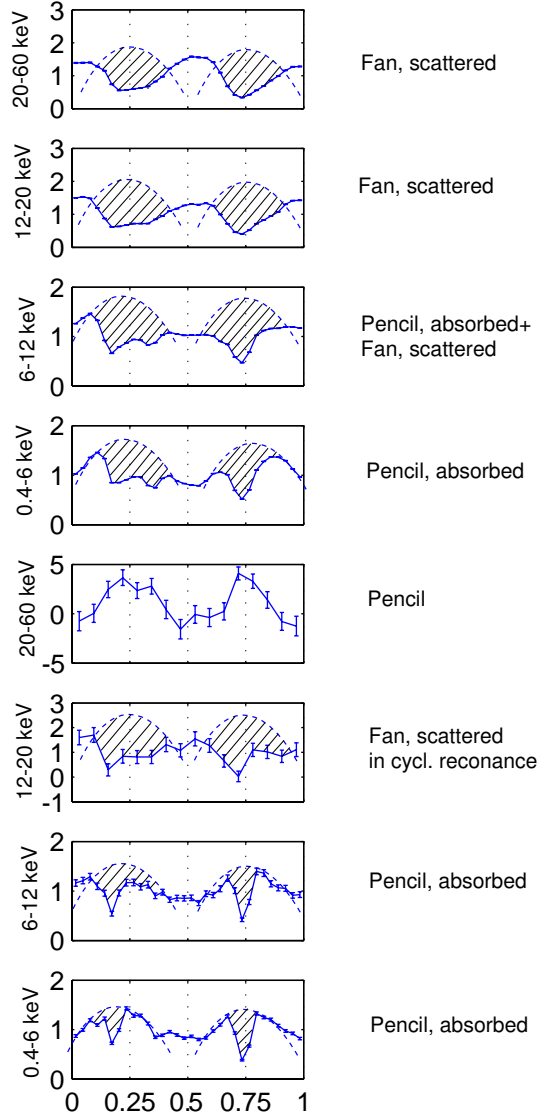}
\caption{Pulse profiles of Vela X-1 as observed by \textit{Suzaku} (in normalized
counts from \citealt{Doroshenko_ea11}) at normal luminosity levels (4 upper panels) and in an `off' state (4 lower panels). The dashed region schematically shows the pencil X-ray beam seen from the magnetic poles of
the neutron star in the `off' state at 20-60 keV (panel 5 from the top) superimposed on the other profiles.}
\label{f:pulses}
\end{figure*} 

The idea that the transition between the two regimes may be triggered by a change in the X-ray beam pattern is supported by observations of Vela X-1. In Fig \ref{f:pulses} we plot the \textit{Suzaku} pulse profiles of Vela
X-1 in different energy bands from the observation by \cite{Doroshenko_ea11}. The top 4 panels show the pulse profiles at normal luminosity levels and the bottom 4 panels show the pulse profiles at different energies within an off-state. The observed change in phase of the 20--60 keV profile in the off-state (at X-ray luminosity $\sim 2.4\times 10^{35}$~erg/s) suggest a disappearance of the fan beam at hard X-ray energies upon the source entering this state.  Instead, a pencil beam formed by thermal O-photons from the two magnetic poles of the neutron star is clearly visible.
This pencil beam is schematically shown as a dashed region superimposed on the other profiles in all panels .

It is of importance here that a cyclotron resonance feature in Vela X-1 is observed
around 20 keV, and that above this energy the scattering cross-section of both O- and X-photons
is essentially equal to the Thomson value. This is why the 20-60 keV pencil beam profile
seen in the `off' state (where the X-ray luminosity $L<L_\dag$) is
strongly scattered by the Thomson-thick accretion flow above the polar caps 
 to form a fan beam at normal luminosity levels. 
Around 20 keV resonance scattering in the cyclotron line above the polar
cap is significant, as suggested by the 12-20 keV profile in the `off' state
(panel number 6 from the top in Fig. \ref{f:pulses}). At lower energies the
absorption in the pencil beam dominates in the `off' state as well as at normal luminosity levels.

\section{Discussion}

We have considered two possible regimes of plasma entering the magnetosphere -- Compton-cooling dominated and radiative cooling dominated subsonic accretion, corresponding to X-ray luminosities
differing by more than an order of magnitude.  
It is essential to realize that the two regimes can be realized for the same density 
of plasma around the magnetosphere. The density $\rho(R_A)$ at the
bottom of the quasi-static shell near the magnetosphere depends on the 
density behind the external shock in the wind at the Bondi capture radius 
$R_B$ as $\rho(R_A)\approx \rho(R_B)(R_B/R_A)^{3/2}\sim \rho_w v_w^{-3}$,
i.e. it is  very sensitive to the stellar wind density and, especially, 
to its velocity.  

On the contrary, the accretion rate onto the neutron star is in our model determined by the ability of the plasma to enter the magnetosphere, i.e. by the plasma settling velocity.
If the radiation energy density is high enough
and the Compton cooling time is shorter than the radiative cooling time of
the plasma, the mean settling plasma velocity $f(u)$ is about 0.3 times the free-fall
velocity. If the radiation density decreases and the radiative cooling time
becomes shorter than the Compton cooling time, the mean settling velocity of the
plasma decreases by a factor of 3, and the mass entry rate onto the magnetosphere
drops by a factor of $\sim 4$ relative to that due to Compton cooling at a given
density $\rho(R_A)$. 

A likely mechanism responsible for the triggering of a transition between the two regimes
is the change of the X-ray emission diagram from the accretion column from 
being essentially a fan beam at high luminosities (large optical thickness of
the accretion column) to a pencil beam (small vertical optical thickness of
the accretion column)  when the mass accretion rate decreases
below $\sim 3\times 10^{15}$~g/s, corresponding to an X-ray luminosity of
$L_\dag=3\times 10^{35}$~erg/s. For the near-magnetosphere density 
corresponding to this X-ray luminosity the
stationary mass entry rate due to radiative cooling is about
$10^{15}$~g/s. Lower mass accretion rates are possible for 
lower external wind densities. 

Clearly, the complicated picture of accretion onto magnetized rotating neutron stars 
is far from being complete. In our model we have taken into account turbulent heating as well as 
Compton and radiative cooling of the plasma near the magnetosphere, 
but ignored the possible effects of magnetic fields
frozen into the plasma, which should be studied separately (see, e.g., resent studies by \cite{Ikhsanov_ea2012}). Magnetic field reconnection,  on the one hand, serves as a source of additional heating of the plasma, but on the other hand, may facilitate plasma entering the magnetosphere
\citep{ElsnerLamb84}. The magnetic reconnection near the magnetosphere
may be responsible for occasional transitions into the 'strong coupling regime', in which
drastic jumps of the neutron star spin period may occur without any change in X-ray luminosity,
as discussed in Paper I. This issue should be further investigated.

\section{Conclusions}

To conclude, quasi-spherical accretion from a stellar wind onto a slowly rotating magnetized
neutron star may proceed in different ways. In high-luminosity
sources with $L_x> 4\times 10^{36}$~erg/s the matter behind the bow shock 
at the Bondi radius cools down rapidly and falls freely towards 
the magnetosphere, forming a shock above the Alfv\'en surface as described by models of Bondi- or supersonic accretion. 

At lower X-ray luminosities a quasi-static atmosphere is bound to be formed above the neutron star
magnetosphere, and accretion proceeds subsonically with an accretion rate
determined by the ability of the plasma to enter the magnetosphere via
instabilities. A model for such subsonic quasi spherical accretion was presented in Paper I. 

At luminosities above $L_\dag\sim 5\times
10^{35}$~erg/s, the plasma cools down via Compton processes and enter the magnetosphere in the
equatorial regions, most favorable for the Rayleigh-Taylor instability
to develop. At X-ray luminosities below $L_\dag\sim 5\times
10^{35}$~erg/s, however, the radiation energy density in the magnetosphere equator 
may be significantly reduced due to a change in the X-ray emission pattern 
from the accretion column. At $L>L_\dag$ the optical depth of the accretion
column is larger than one and radiation is scattered mostly in the
lateral direction forming a fan beam, at $L_x<L_\dag$ the vertical Thomson optical depth in the
accretion column becomes smaller than one, and a pencil-beam is formed. 
The pencil beam illuminates the magnetosphere cusp region, where plasma
entry is hampered by large curvature of the magnetic field lines. Still, the
plasma continues to find its way onto the magnetic field lines in the equatorial 
region due to radiative plasma cooling, which is determined by the plasma
density independently of the radiation energy density. The mass accretion 
rate due to radiative cooling is by several times smaller than due to 
Compton cooling with the same plasma density above the Alfv\'en surface. 

We identify the two subsonic regimes; the Compton cooling dominated and the radiative cooling dominated, respectively, with observed `normal'  luminosity levels and the so called `off' states in some slowly
rotating X-ray pulsars, e.g. Vela X-1, GX 301-2, 4U1907+05. Our proposed scenario is supported by the observed
change in the hard X-ray pulse profile in the off state of Vela X-1 and could be further checked against observations of the behaviour of hard X-ray pulse profiles during a transition into or out of an off-state.

It also can not be excluded that  
the phenomenon of Supergiant Fast X-ray Transients (SFXTs) (see \citealt{Sidoli11} for 
a recent summary and review) can similarly be related to transitions between 
different regimes of plasma cooling in a quasi-spherical
shell around a slowly rotating magnetized neutron star. The quiescent states of SFXTs with 
stable low-luminosity accretion with $L_{rad}\sim 10^{34}$~erg/s may be  
controlled by thermal plasma cooling, while the unstable X-ray flares may be triggered 
by a density increase above the magnetosphere
leading to an increase in the lateral X-ray emission from the accretion
column and a transition to the Compton-cooling dominated regime. Our model predicts the corresponding 
change in hard X-ray pulse profile to occur during the transition from the quiescent state 
to the flaring state, which can be checked by dedicated observations.

\section{Acknowledgements}
The authors thank Dr. V. Doroshenko for providing data on the pulse profiles of
Vela X-1. The remaining data presented in this paper were obtained through the High Energy Astrophysics Science Archive Research Center (HEASARC) Online Service, provided by NASA/Goddard Space Flight Center. The work by NSh and KP was supported by RFBR grants 12-02-00186a and
10-02-00599a. LH was supported by a grant from the Wenner-Gren foundations.

\end{document}